\documentclass[aps, prl, amsmath, floats,floatfix, twocolumn, superscriptaddress]{revtex4-1}
 
\usepackage[normalem]{ulem}
\usepackage{amssymb}
\usepackage{amsmath}
\usepackage{verbatim}
\usepackage{mathrsfs}
\usepackage{amsfonts}
\usepackage{latexsym}
\usepackage{epsfig}
\usepackage{color}
\usepackage{graphicx,subfigure}
\usepackage{units}
\usepackage{slashbox}
\usepackage{envmath}
\usepackage{natbib}
\usepackage{hyperref}
\usepackage[utf8]{inputenc}
\usepackage{bm}

\begin{document}


\definecolor{orange}{rgb}{0.9,0.45,0}

\newcommand{\re}{\mbox{Re}}
\newcommand{\im}{\mbox{Im}}
\newcommand{\ch}[1]{\textcolor{red}{CH: #1}}
\newcommand{\er}[1]{\textcolor{cyan}{[\bf ER: #1]}}

\def\CovDev{D}
\def\Res{{\mathcal R}}
\def\Gammaflat{\hat \Gamma}
\def\metricflat{\hat \gamma}
\def\Dflat{\hat {\mathcal D}}
\def\part_n{\partial_\perp}

\def\Lie{\mathcal{L}}
\def\A{\mathcal{X}}
\def\Aphi{\A_{\phi}}
\def\hAphi{\hat{\A}_{\phi}}
\def\E{\mathcal{E}}
\def\Ham{\mathcal{H}}
\def\M{\mathcal{M}}
\def\R{\mathcal{R}}
\def\p{\partial}

\def\hg{\hat{\gamma}}
\def\hA{\hat{A}}
\def\hD{\hat{D}}
\def\hE{\hat{E}}
\def\hR{\hat{R}}
\def\hcA{\hat{\mathcal{A}}}
\def\hDelt{\hat{\triangle}}

\def\be{\begin{equation}}
\def\ee{\end{equation}}

\renewcommand{\t}{\times}

\long\def\symbolfootnote[#1]#2{\begingroup%
\def\thefootnote{\fnsymbol{footnote}}\footnote[#1]{#2}\endgroup}

\title{(H)airborne:\\ two spinning black holes balanced by their synchronised scalar hair}

\author{
Carlos A. R. Herdeiro,
Eugen Radu
}

 \affiliation{Departamento de Matem\'atica da Universidade de Aveiro and CIDMA,
Campus de Santiago, 3810-183 Aveiro, Portugal}
%


\date{May 2023}

\begin{abstract}
General Relativity minimally coupled to a massive, free, complex scalar field, is shown to allow asymptotically flat solutions, non-singular on and outside the event horizon, describing two spinning black holes (2sBHs) in equilibrium, with  co-axial, aligned angular momenta.  The 2sBHs configurations bifurcate from solutions describing dipolar spinning boson stars. The BHs emerge at equilibrium points diagnosed by a test particle analysis and illustrated by a Newtonian analogue. The individual BH ``charges" are mass and angular momentum only.   Equilibrium is due to the scalar \textit{environment}, acting as a (compact) dipolar field, providing a lift against their mutual attraction, making the 2sBHs (h)airborne. We explore the 2sBHs domain of solutions and its main features.
\end{abstract}




\maketitle
{\bf Introduction.}
The equilibrium 2-body problem provides a theoretical window into the non-linear interactions between compact objects, in particular black holes (BHs), in General Relativity (GR). 

Vacuum BHs -- the \textit{gravitational atoms} -- have only mass and angular momentum, as  macroscopic degrees of freedom~\cite{Carter:1971zc,Robinson:1975bv,Chrusciel:2012jk}, measured at their horizon, $(M_H,J_H)$. It turns out that such ``atoms"  
cannot be balanced in vacuum GR, without introducing naked singularities~\cite{Costa:2009wj,Neugebauer:2011qb,Hennig:2011fp}. 
Placing them in an external \textit{global} gravitational field, on the other hand, can indeed balance gravitational atoms \textit{locally}~\cite{Astorino:2021dju}, but at the cost of losing asymptotic flatness, in fact creating asymptotic singularities~\footnote{The latter pathology is circumvented by replacing the external field by a positive cosmological constant~\cite{Dias:2023rde}; these solutions are asymptotically de Sitter, rather than asymptotically flat.}. 
This suggests, however, that the right \textit{smooth, local environment} could balance gravitational atoms~\footnote{Balance can also be achieved by endowing the individual BHs with extra macroscopic repulsive degrees of freedom -  gauge charges - $Q_H$. The simplest such balanced configuration is the Majumdar-Papapetrou~\cite{Majumdar:1947eu,Papapetrou:1948jw} solution of electrovacuum. Here we focus on neutral BHs, with  $(M_H,J_H)$ as the only charges measured at the horizon.}. 
This is corroborated by a recent example~\cite{Herdeiro:2023mpt}, wherein, however, the environment was made up of unphysical ``matter".

In this letter we show that equilibrium \textit{gravitational molecules} - two BHs with only $(M_H,J_H)$ individual degrees of freedom and asymptotically flat -, exist in a smooth local environment, as exact solutions of GR minimally coupled to simple and physical ``matter". More generically, we unveil a mechanism, and lay out a methodology, to construct other families of such non-linear molecules, prone to different applications.

{\bf The model and the environment.} 
Consider the Einstein-(massive, complex) scalar model described by the action $\mathcal{S}=(4\pi)^{-1}\int d^4 x
\sqrt{-g}\mathcal{L}$, with~\footnote{We use geometrized units where $G=1=c$.} 
\begin{eqnarray}
\label{action}
 \mathcal{L}=
\frac{R}{4}
   - g^{\alpha\beta}
	\Phi_{, \, \alpha}^* \Phi_{, \, \beta}    -\mu^2 \Phi^* \Phi	\ ,
\end{eqnarray}
where $R$ is the Ricci scalar, $g_{\alpha\beta}$ the
spacetime metric, with determinant $g$, 
$\Phi$ is a complex scalar field with mass $\mu$, 
and
the asterisk denotes complex conjugation. 

Model~\eqref{action}, for appropriate ranges of $\mu$, is a popular fuzzy dark matter model, to which different scalar self-interactions can be added~\cite{Suarez:2013iw,Hui:2016ltb}.
One class of solutions are scalar (mini-)boson stars (BSs)~\cite{Schunck:2003kk}, a family of self-gravitating solitons, often described as macroscopic Bose-Einstein condensates~\cite{Liebling:2012fv}, that can have a multipolar structure, akin to hydrogenic orbitals~\cite{Herdeiro:2020kvf}. Here we focus on dipolar spinning boson stars (DsBSs)~\cite{Kleihaus:2007vk,Kunz:2019bhm} - see also~\cite{Yoshida:1997nd,Cunha:2022tvk} for the static case - with the metric 
\begin{equation}
\label{ansatz}
ds^2=-e^{2F_0} dt^2+e^{2F_1}\left(dr^2  + dz^2\right)
+e^{2F_2}\rho^2 (d\varphi-W dt)^2  , 
\end{equation} 
where $\{\rho,z,\varphi \}$ 
are cylindrical coordinates and $F_i,W$ are functions of $(\rho,z)$, $i=0,1,2$.
The geometry of DsBSs is $\mathbb{Z}_2$-even,
$F_{i}(\rho,-z)=F_{i}(\rho, z)$,
$W(\rho,-z)=W(\rho, z)$, with an equatorial plane at $z=0$. On the other hand, the scalar field, 
$\Phi=\phi(\rho,z) e^{i(m \varphi-\omega t)}$, is $\mathbb{Z}_2$-odd, $\phi(\rho,z)=- \phi(\rho,-z)$,  where $\omega>0$  and 
$m\in \mathbb{Z}$ are the frequency and azimuthal harmonic index. The $\pi$ phase difference between the north and south hemispheres yields a repulsive scalar interaction between the two constituents of the DsBSs~\cite{Palenzuela:2006wp,Cunha:2022tvk}. DsBSs can be interpreted as two  \textit{individual} spinning BSs balanced by their short (long) range scalar repulsion (gravitational attraction). The domain of existence of the DsBSs is shown in the inset of Fig.~\ref{fig2} in an ADM mass $vs.$ frequency diagram~\footnote{In the following, dimensionful quantities, such as frequency $\omega$ and mass $M$ are presented in units of $\mu$.}. The stars exist for $0.6835\lesssim \omega\leqslant 1$ and the diagram shows the characteristic spiral shape of other bosonic star solutions.

\begin{figure}[h!]
\begin{center}
\includegraphics[width=0.45\textwidth]{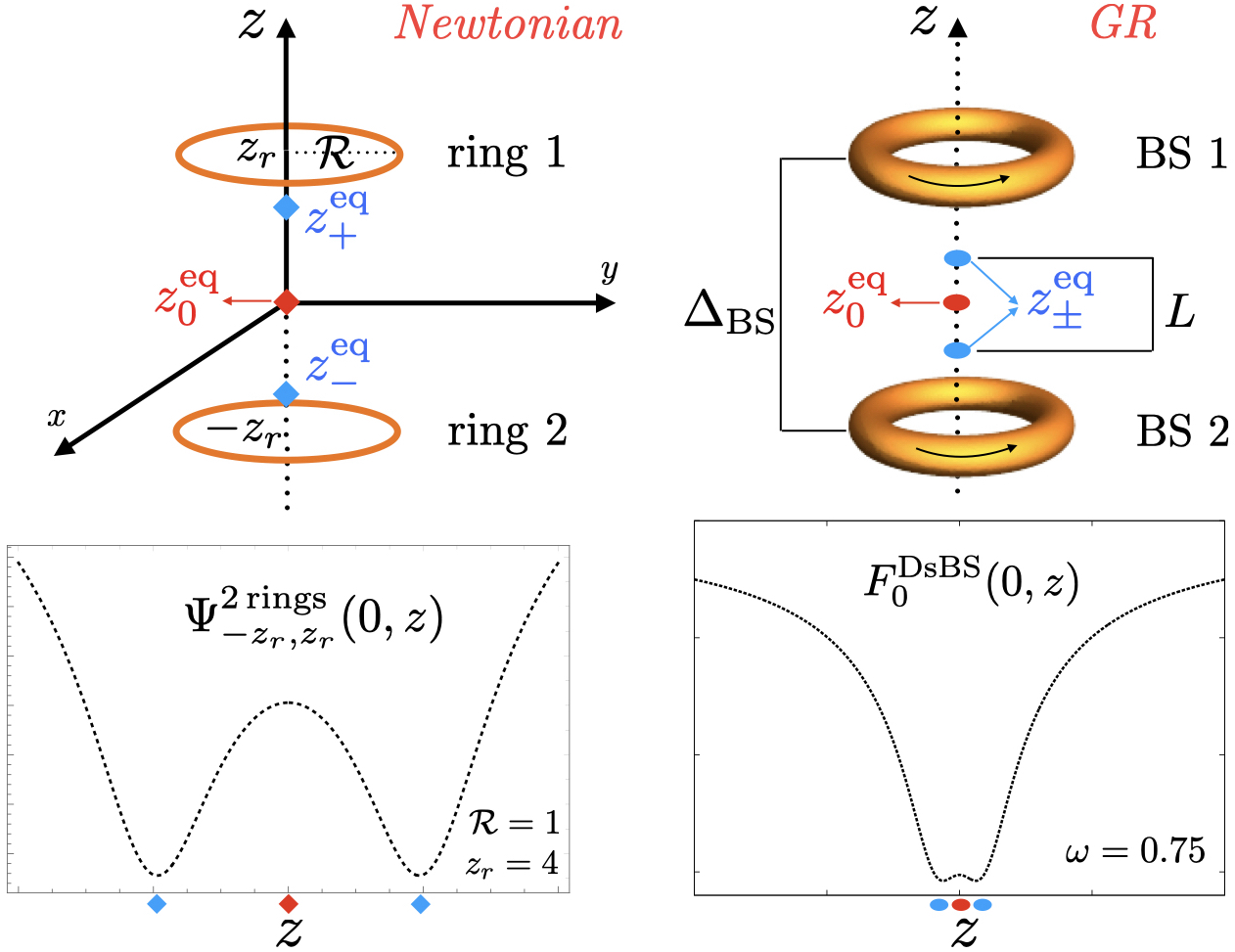}
\caption{\small  Equilibrium points in: (left) a Newtonian 2-rings system (diamonds) and the corresponding gravitational potential along the $z$-axis; (right) the GR DsBSs system (disks) and a corresponding (illustrative) metric $F_0$ function. $L$ ($\Delta_{\rm BS}$) will denote the proper distance between the horizons (BSs).}
\label{fig1}
\end{center}
\end{figure}

\begin{figure}[h!]
\begin{center}
\includegraphics[width=0.45\textwidth]{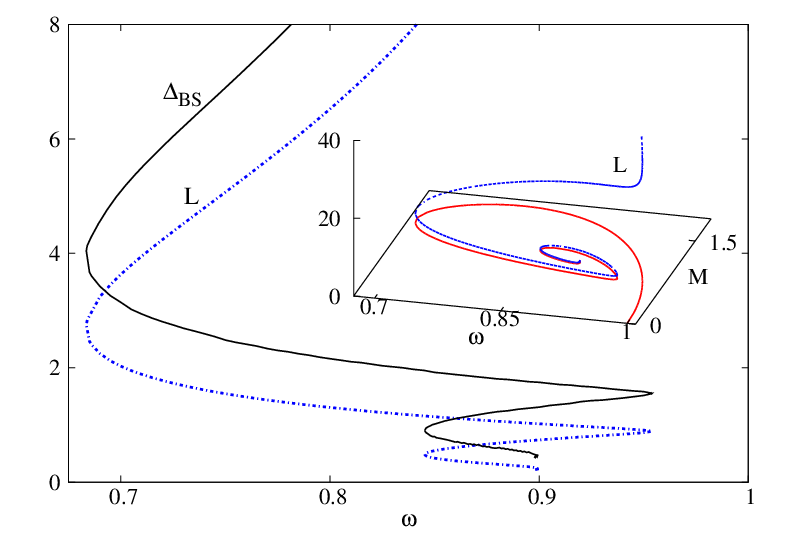}
\caption{\small Domain of existence of DsBSs (inset - red line) and the distance between the two constituents $\Delta_{\rm BS}$ (main panel) $vs.$ $\omega$. $\Delta_{\rm BS}$ is the proper distance along the $z$-axis between the extrema of $\Phi$, attained, however, for $\rho\neq 0$, due to the toroidal morphology. $L$ is the proper distance between $z_\pm^{\rm eq}$.}
\label{fig2}
\end{center}
\end{figure}

{\bf Bifurcation points.}
Individual spinning mini-BSs are mass tori in GR~\cite{Schunck:1996he,Yoshida:1997qf}. To gain intuition, consider a mass torus in \textit{Newtonian} gravity. The simplest case is  an infinitely thin torus: a ring of radius $\mathcal{R}$ and constant mass density $\chi$. In a cylindrical chart $(\rho,z,\phi)$ on $\mathbb{R}^3$~\footnote{I.e.~the spatial metric~\eqref{ansatz} with $F_i=W=0$.}, take the ring on the $z=z_r\geqslant 0$ plane and centred at $\rho=0$; its gravitational potential at $(\rho,z,\phi)$ is 
\begin{equation}
\Psi_{z_r}^{\rm ring}=-\int_0^{2\pi}\frac{\chi \mathcal{R} d\varphi}{\sqrt{\rho^2-2\rho \mathcal{R} \cos (\phi-\varphi)+\mathcal{R}^2+(z-z_r) ^2}} \, .
\end{equation}
The gravitational force ${\bf F}=-\nabla \Psi$ has a $z$-component always directed towards the plane of the ring (for particles outside that plane). Its magnitude is not monotonic. Along $\rho=0$, it attains a maximum at a critical distance
$
|\tilde{z}-\tilde{z}_r|=\tilde{z}^{\rm crit}\equiv {1}/{\sqrt{2}}$,
where $\tilde{z}\equiv z/\mathcal{R}$. This turns out to determine the equilibrium points of a 2-rings system, a Newtonian analogue to the DsBSs solutions - Fig.~\ref{fig1}.

Consider two parallel thin rings, both with radius $\mathcal{R}$, symmetrically placed with centres at $\rho=0$, on the planes $z=\pm z_r$ - Fig.~\ref{fig1} (top left). 
The corresponding Newtonian potential is $
\Psi^{\rm 2 \, rings}_{-z_r,z_r}=\Psi_{-z_r}^{\rm ring}+\Psi_{z_r}^{\rm ring}$. The number of equilibrium points on the $z$-axis now depends on the dimensionless ratio $\tilde{z}_r$:
for $\tilde{z}_r<\tilde{z}_r^{\rm crit}$, there is a single equilibrium point, at $z_0^{\rm eq}=0$. This is stable against vertical displacements, since a test particle displaced from the origin is attracted to the \textit{furthest} ring; 
for $\tilde{z}_r> \tilde{z}_r^{\rm crit}$, 2 new equilibrium points emerge, $0<|z_{\pm}^{\rm eq}|<|z_r|$. They bifurcate from $z_0^{\rm eq}=0$ and are symmetric with respect to $z=0$~\footnote{These are determined by $(z_{\pm}^{\rm eq}-z_r)[\mathcal{R}^2+(z_{\pm}^{\rm eq}+z_r)^2]^{3/2}+(z_{\pm}^{\rm eq}+z_r)[\mathcal{R}^2+(z_{\pm}^{\rm eq}-z_r)^2]^{3/2}=0$.} - see an illustrative $\Psi$ in Fig.~\ref{fig1} (bottom left). They are stable against vertical displacements, whereas $z_0^{\rm eq}=0$ now becomes unstable. In the limit $\tilde{z}_r\gg 1$, the new equilibrium points tend to the centre of each of the rings,  $z_{\pm}^{\rm eq}\rightarrow \pm z_r$. The upshot is that the emergence of the new equilibrium points relies on the existence of a maximum of the vertical force of each ring. 

Let us now turn to the (timelike) equilibrium points on the DsBSs geometry, $cf.$ eq.~\eqref{ansatz}, of the form $(\rho,z,\varphi)=(0,z^{\rm eq},\varphi_0);(\dot \rho,\dot z,\dot \varphi)=0=(\ddot \rho,\ddot z,\ddot \varphi)$, where derivatives are with respect to proper time. The geodesic equations, $\ddot x^{\alpha}+\Gamma^\alpha_{\beta\gamma}\dot x^\beta \dot x^\gamma=0$, yield the Christoffel symbols conditions $\Gamma^\rho_{tt}=
\Gamma^z_{tt}=\Gamma^\varphi_{tt}=0$, at the equilibrium point. Considering the geometry~\eqref{ansatz}, the last condition is obeyed and the first follows from smoothness at the axis of the metric functions. The remaining condition ($\Gamma^z_{tt}=0$) yields $\partial_z F_0=0$ at equilibrium points. We scanned the parameter space of DsBSs~\cite{Kunz:2019bhm} - Fig.~\ref{fig2} - and observed always \textit{three} equilibrium points, $z_0^{\rm eq}=0$, $z_\pm^{\rm eq}\neq 0$, as in the Newtonian 2-rings system for $\tilde{z}_r> \tilde{z}_r^{\rm crit}$ - Fig.~\ref{fig1} (right).  The fact that $\Delta_{\rm BS}$ (Fig.~\ref{fig2}) never nears zero may explain the absence of DsBSs with a single equilibrium point.

Static equilibrium points for a timelike test particle in a BH/soliton spacetime suggest (but do not guarantee) a bifurcation towards a new family of solutions, wherein the test particle is promoted to a back-reacting object. A familiar example is the extremal Reissner-Nordstr\"om BH which bifurcates into the Majumdar-Papapetrou multi-BH spacetimes, which can be diagnosed by the equilibrium points of (charged) geodesics around the former. Another example, closer to our discussion, are spinning BSs, that have an equilibrium point at their centre. A spinning BH with horizon angular velocity $\Omega_H$ can be added therein under the \textit{synchronisation condition}~\cite{Herdeiro:2014goa} 
\begin{equation}
w=m\Omega_H \ .
\label{sync}
\end{equation}
The resulting \textit{hairy} BHs interpolate between spinning BSs and the Kerr solution~\cite{Kerr:1963ud}. On the other hand, the bifurcation from the latter end is diagnosed by a different equilibrium: the existence of a zero mode of a test bosonic field, at the threshold of superradiance~\cite{Hod:2012px}.

We shall now show that \textit{two} BHs can be grown from the equilibrium points of the DsBSs. They can either \textit{both} emerge from $z^{\rm eq}_0$ or \textit{one from each} of the two $z^{\rm eq}_\pm$ points. Equilibrium between each horizon and the scalar environment requires condition~\eqref{sync}. Equilibrium between the two horizons relies on the extra opposing force of the two DsBS constituents, themselves readjusting their distance $\Delta_{\rm BS}$ as the BHs are grown. One anticipates that equilibrium requires BHs carrying a small enough fraction of the total mass. This is confirmed below. A complementary perspective is that each horizon-BS pair is a BH with synchronised scalar hair~\cite{Herdeiro:2014goa}, with the repulsive force due to the $\pi$-phase difference between the two bosonic clouds   balancing the system.

{\bf Generalized Weyl framework.}
Axially symmetric, stationary, multi-BH solutions in electrovacuum can be analytically constructed as Weyl solutions~\cite{Weyl:1917gp,Emparan:2001wk}. At the heart of this construction is the notion of \textit{rod structure}~\cite{Harmark:2004rm}. In~\cite{Herdeiro:2023mpt},
a generalized Weyl construction for numerical implementation  in models wherein integrability is lost, such as~$\eqref{action}$,  was introduced, still based on a rod-structure. We shall follow this construction, generalizing it for stationary (rather than static) solutions. 

In our case, the rod structure introduces 2 positive parameters: $\Delta z$, (roughly) measuring the distance between horizons and $z_H$, determining the horizon(s) size; it reads:
$(i)$ a $1^{st}$ semi-infinite spacelike rod, $[-\infty,-z_H]$, in the $\varphi$-direction;
$(ii)$
a $1^{st}$ (finite) timelike rod, $[-z_H, -\Delta z/2]$;
$(iii)$
a finite spacelike rod, $[-\Delta z/2, \Delta z/2]$;
$(iv)$
a $2^{nd}$ (finite) timelike rod $[\Delta z/2 , z_H]$;
$(v)$
a $2^{nd}$ semi-infinite spacelike rod, $[z_H,\infty]$.

This rod structure results in a set of boundary conditions on the $z$-axis, $\rho=0$, conveniently expressed in terms of $f_0\equiv e^{2F_0},f_1\equiv e^{2F_1}, f_2\equiv \rho^2e^{2F_2}$. These are as follows. For $m\neq 0$, $\phi=0$. For the timelike rods, corresponding to the BH horizons, 
$ f_0=
\partial_\rho f_1=
\partial_\rho f_2=0,
  W=\Omega_H,$ together
 with $\lim_{\rho\to 0}\rho^2 f_1/f_0=$const.
and $\partial_\rho W=0$.
The horizon angular velocity of the two BHs is the same, 
$\Omega_H=W (-z_H \leqslant z \leqslant -\Delta z/2)=W (\Delta z/2\leqslant z \leqslant z_H)$;
 the two BHs are {\it corotating}. As with the single BH case \cite{Herdeiro:2014goa}, assuming the existence of a power series expansion 
of the solutions near a horizon results in~\eqref{sync}. 
Each horizon has a spherical topology,
but deviating from a 
round sphere. They have the same area
$A_H= 2\pi \int_{ \Delta z/2 }^{ z_{H} }dz \sqrt{f_1(0,z)f_2(0,z) }$ and 
Hawking temperature,
$T_H=\lim_{\rho\to 0} \sqrt{{f_{0}(\rho,z)}/[\rho^2 f_{1}(\rho,z)}]/2\pi$.
Thus the horizons are in thermodynamical equilibrium. The total entropy is twice that of a single BH, $
 S=A_H/2$.

On a spacelike rod we impose (again at $\rho=0)$  $\partial_\rho  f_0=\partial_\rho f_1=
  f_2=
\partial_\rho W=0$;
$\lim_{\rho\to 0}\rho^2 f_1/f_2=1$, additionally, 
imposes 
the absence of conical singularities.
The proper
length of the finite $\varphi$-rod measures the inter-horizons distance, 
$ L=\int_{-\Delta z/2}^{\Delta z/2} dz\sqrt{f_1(0,z)}$ - Fig~\ref{fig1} (right).

 Flat spacetime is approached for large $(\rho,|z|)$,   
wherein $\phi\rightarrow 0$.
The ADM mass $M$  and angular momentum $J$ 
 can be read off from the asymptotic metric components: 
$
-g_{tt}
\simeq 1-{2  M}/{ \sqrt{\rho^2+z^2}}$, $
 g_{\varphi t} 
\simeq 
 -{2 J \rho^2}/{ \left(\rho^2+z^2 \right)^{3/2} }$.
As usual in (asymptotically flat) BH mechanics, 
the temperature, entropy and the global charges
are related by a Smarr mass formula \cite{Townsend:1997ku},
$
M=2 T_H S  +2\Omega_H (J-m Q)+ M_\Phi$,
with
$M_\Phi$
 the energy $outside$ the BHs stored in the scalar field. $Q$ is the conserved  Noether charge associated to the global $U(1)$ symmetry of~\eqref{action}~\footnote{Physical quantities are also connected via the first law, which, for balanced configurations, takes the usual form $dM=T_H dS +\Omega_H dJ$.}.
The total mass is then $M=M_H+M_\Phi$, with $M_H$ the (total) mass of the horizons, that can be computed via Komar integrals~\cite{Herdeiro:2015gia}.

Finally,  the vacuum, co-rotating
double Kerr solution~\cite{KRAMER1980259} can be described by the above formalism, but with rather involved expressions  - see $e.g.$ \cite{Costa:2009wj,Cabrera-Munguia:2017dol,Manko:2017avt}.

{\bf Numerics.} 
We have solved numerically the  set of five coupled
non-linear elliptic partial differential equations resulting from~\eqref{action}, with the ansatz~\eqref{ansatz},  subject to the above boundary
conditions. For $\Phi$ we take $m=1$~\footnote{We have found solutions with $m=2$ as well, not reported here.}. The input parameters are $\{\Omega_H \ {\rm or} \ w; \Delta z,z_H\}$.  The balanced two spinning BHs (2sBHs) are constructed by scanning the domain of existence of general unbalanced solutions at constant $\Omega_H$,
and varying $z_H$ and $\Delta z$. Additionally fixing (say)  $z_H$, regular configurations may exist for some discrete values of 
$\Delta z$ only~\footnote{The accurate construction of solutions with 
large frequencies $\omega \gtrsim 0.9$ remains a numerical challenge.}. Details on the approach, in particular on the coordinates better suited for numerics, can be found in~\cite{Herdeiro:2023mpt}.
All solutions reported here are free of conical singularities, and no pathologies were observed
on and outside the event horizon.
Typical numerical errors
are $\sim 10^{-3}$.

{\bf Results.} 
DsBSs have no timelike rods: $z_H=\Delta z/2$. $\omega$ is the only input parameter ($\Delta z$ is arbitrary).  Taking $\Delta z=0,z_H\neq 0$ adds a \textit{single} BH horizon  at $z_0^{\rm eq}=0$ for~\textit{any} DsBS along the spiral in Fig~\ref{fig2}, yielding a family of BHs with 
 parity-odd  synchronised hair, studied in~\cite{Kunz:2019bhm}.

\begin{figure}[h!]
\begin{center}
\includegraphics[width=0.45\textwidth]{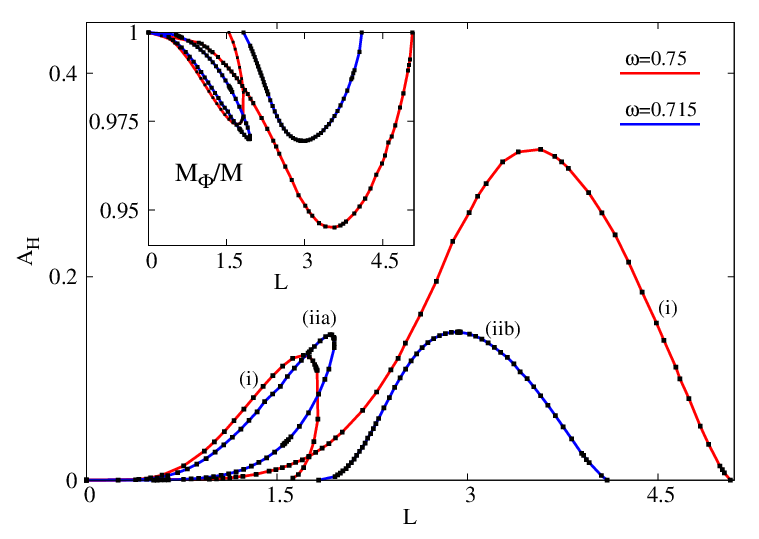}
\caption{\small Horizon area $A_H$ $vs.$ inter-horizon distance $L$ for qualitatively different sequences of (balanced) 2sBHs. The inset shows the mass fraction stored in the scalar field. }
\label{fig3}
\end{center}
\end{figure}

\textit{Any} DsBS appears to \textit{also} bifurcate into a 2sBH configuration, obeying (\ref{sync}).
%
As for the single BH case \cite{Kunz:2019bhm}, increasing the horizon size - via $z_H$ -, 
2BHs emerge from the seed DsBS at $z_0^{\rm eq}=0$. Along the branch with the corresponding fixed frequency $\omega=\Omega_H$, 
 the two BHs initially increase in size ($A_H$). They depart from the equatorial plane - $L$ grows - becoming (h)airborne. The sequence of solutions can terminate in \textit{two} possible ways. Away from the minimal frequency, $\omega\gtrsim 0.735$, $L$ never vanishes again and the sequence ends on the  $z_\pm^{\rm eq}\neq 0$ equilibrium points of the \textit{same} DsBS - this is illustrated by two sequences labelled $(i)$ in Fig.~\ref{fig3}. Close to the minimal frequency, $0.6835\lesssim\omega\lesssim 0.735$, 
$L$ does vanish again and the sequence ends on the  $z_0^{\rm eq}=0 $ equilibrium point of \textit{another} DsBS (with the same $\omega$ - Fig.~\ref{fig2}) - this is labelled  sequence $(iia)$ in Fig.~\ref{fig3}. Additionally, these very same two DsBSs can also be connected by a sequence departing from the  $z_\pm^{\rm eq}\neq 0$ equilibrium points of one and arriving at the analogous points of the other; here $L$ never vanishes along the whole sequence,  labelled $(iib)$ in Fig.~\ref{fig3}.

\begin{figure}[h!]
\begin{center}
\includegraphics[width=0.45\textwidth]{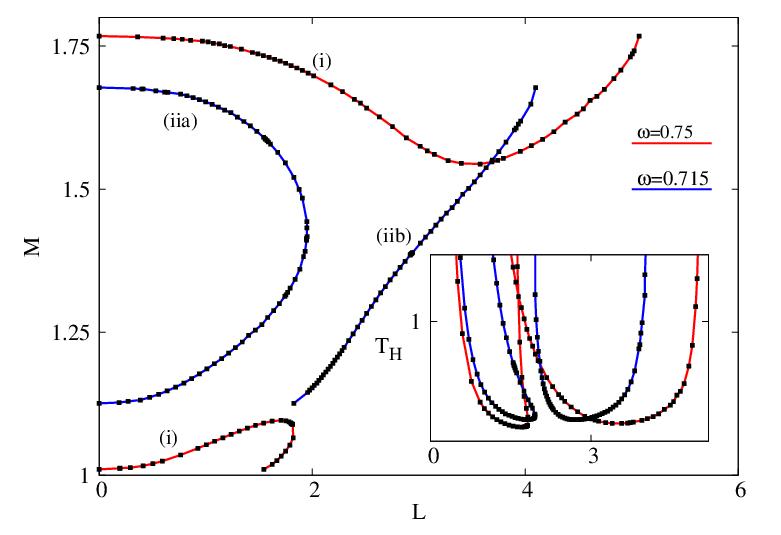}
\caption{\small Mass $M$ $vs.$ inter-horizon distance $L$ for the same sequences as in Fig.~\ref{fig3}. The inset shows the BH temperature.}
\label{fig4}
\end{center}
\end{figure}

Fig.~\ref{fig3} (inset) also reveals that the horizon mass is never larger than a few percent - in our scanning we found a maximum of $\sim 8\%$ total horizon mass (twice that of an individual BH) along a type $(i)$ sequence, seeded on the upper DsBSs branch. The largest BHs are also the ones with the lowest Hawking temperature - Fig.~\ref{fig4} (inset). The main panel of this figure also makes clear that the two DsBSs connecting sequences $(iia)/(iib)$ are different, since they have different masses. 

\begin{figure}[t!]
\begin{center}
\includegraphics[width=0.45\textwidth]{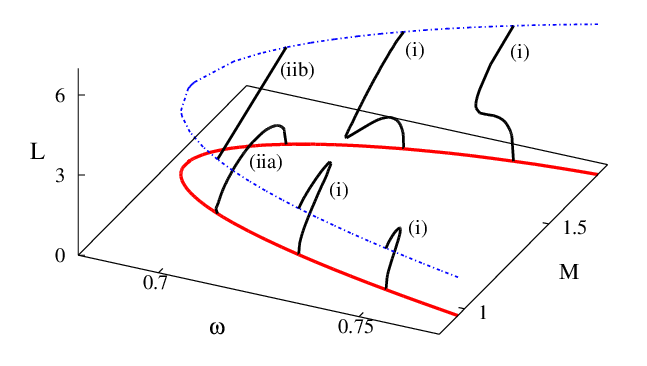}
\includegraphics[width=0.4\textwidth]{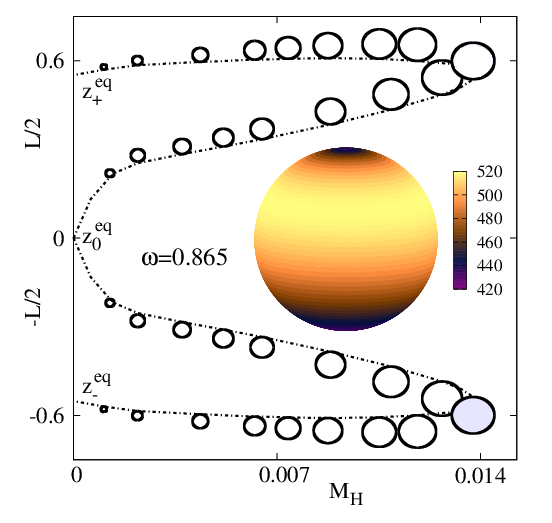}
\caption{\small (Top) Illustrative sequences of 2sBHs branching off from the DsBHs. (Bottom)  Horizon embedding superimposed on an $(M_H,L)$-diagram. The heatmap shows the horizon geometry Ricci scalar for the maximal area solution (the south one - in blue), which varies by $\sim$20\% and is $\mathbb{Z}_2$-asymmetric, maximized in the northern hemisphere.}
\label{fig5}
\end{center}
\end{figure}

Fig.~\ref{fig5} (top panel) provides a different illustration of the sequences $(i),(iia)$ and $(iib)$. It zooms on the lower frequency end of the $M$-$\omega$ domain of existence of DsBSs - inset of Fig.~\ref{fig2} -, since the transition between sequences $(i)$ and $(ii)$ occurs in this frequency region. Adding $L$ as a third dimension, makes transparent the sequences of 2sBHs (black solid lines), branching off from the DsBSs at $z_0^{\rm eq}=0$ - from the red solid line -  or at $z_\pm^{\rm eq}\neq 0$ - from the blue dotted line, which yields $L$. %
From the drawn sequences one can extrapolate the surface describing the domain of existence of the 2sBHs: two tendentially vertical (curved) surfaces for the larger $\omega$ that join together and become two tendentially horizontal (curved) surfaces for the smaller $\omega$. Additionally, one can conceive another surface, with $L=0$, bounded by the DsBS line, describing the \textit{single} BH solutions found in~\cite{Kunz:2019bhm}.

Fig.~\ref{fig5} (bottom panel) shows a sequence of type $(i)$ in a horizon mass $M_H$ $vs.$ distance $L$ diagram, for a different (larger) frequency. It also superimposes the horizon shape and size, computed as an isometric embedding in $\mathbb{E}^3$~\cite{Smarr:1973zz}. One can appreciate the growth of the horizon size from both the $z_0^{\rm eq}=0$ and the $z_\pm^{\rm eq}\neq 0$ branching points. The deviation from sphericity of the horizons is small; the individual horizons also  possess a small $\mathbb{Z}_2$ asymmetry, better seen in the heat map inset.

{\bf Discussion.}
The asymptotically flat balanced 2sBHs described  rely on two ingredients: the synchronisation condition~\eqref{sync}, allowing equilibrium between a bosonic environment and a horizon; a bosonic environment whence the BHs bifurcate with multiple (three) equilibrium points. 
These ingredients are also present - possibly with a slightly different rationale - in other models leading to natural generalization of these balanced 2sBHs solutions. Adding scalar self-interactions to~\eqref{action} allows a variety of models~\cite{Schunck:2003kk}. Judging by the healing properties seen in other contexts~\cite{DiGiovanni:2020ror,Siemonsen:2020hcg,Sanchis-Gual:2021phr}, self-interactions may improve the dynamical properties of these 2sBHs systems, an interesting (but challenging) problem. Similar 2sBH configurations may also exist for Proca fields, for which BHs with synchronised hair are also known~\cite{Santos:2020pmh}, but (currently) not dipolar Proca stars, static or spinning.

A detailed study of the physical and mathematical properties of these balanced 2sBHs systems is in order and will be presented somewhere else. Phenomenologically, since Eq.~\eqref{ansatz} describes a dark matter model, one may entertain the question if there could be realistic dark matter distributions that would bringing the mutual attraction of a \textit{dynamical} BH binary to a halt.

\newpage

{\bf Acknowledgements.}
We thank Pedro Cunha and Denjoe O'Connor for valuable  discussions. This work is supported by the Center for
Research and Development in Mathematics and Applications (CIDMA) through the Portuguese Foundation for Science and Technology (FCT – Fundação para a Ciência e a Tecnologia), references UIDB/04106/2020 and UIDP/04106/2020. The authors acknowledge support from the projects CERN/FIS-PAR/0027/2019, PTDC/FIS-AST/3041/2020, as well as CERN/FIS-PAR/0024/2021 and 2022.04560.PTDC. This work has further been supported by the European Union’s Horizon 2020 research and innovation (RISE) programme H2020-MSCA-RISE-2017 Grant No. FunFiCO-777740 and by the European Horizon Europe staff exchange (SE) programme HORIZON-MSCA-2021-SE-01 Grant No. NewFunFiCO-101086251. Computations were performed at the ARGUS cluster at Aveiro University.

\bibliography{num-rel2}


 
\end{document}